\documentclass[aps,prl,preprint,showpacs,groupedaddress]{revtex4}
\usepackage{graphics}
\usepackage{multirow}
\usepackage{epsfig}

\begin{document}

\title{Measurement of Land\'{e} g factor of $5\mathrm{D}_{5/2}$ state of BaII with a single trapped ion}

\author{N. Kurz}
\email[]{nkurz129@u.washington.edu}
\homepage[Group homepage\hspace{5pt}]{http://depts.washington.edu/qcomp/}
\author{M. R. Dietrich}
\altaffiliation{Current address: Physics Division, Argonne National Laboratory, Argonne, IL 60439}
\author{Gang Shu}
\author{T. Noel}
\author{B. B. Blinov}
\affiliation{Department of Physics, University of Washington, Seattle, WA 98195}

\date{\today}

\begin{abstract}
We present the first terrestrial measurement of the Land\'{e} g factor of the $5\mathrm{D}_{5/2}$ state of singly ionized barium.  Measurements were performed on single Doppler-cooled $^{138}\mathrm{Ba}^+$ ions in a linear Paul trap.  A frequency-stabilized fiber laser with nominal wavelength 1.762 $\mu$m was scanned across the $6\mathrm{S}_{1/2}\leftrightarrow 5\mathrm{D}_{5/2}$ transition to spectroscopically resolve transitions between Zeeman sublevels of the ground and excited states.  From the relative positions of the four narrow transitions observed at several different values for the applied magnetic field, we find a value of $1.2020\pm0.0005$ for $g_{5\mathrm{D}_{5/2}}$.
\end{abstract}



\maketitle

Single trapped ions continually prove themselves to be a valuable physical system for high-precision measurements motivated by study of the variation of fundamental constants\cite{Prestage, Fortier, Dzuba}, the construction of frequency standards \cite{standards, clocks} and for tests of violations of fundamental symmetries and physics beyond the standard model \cite{FundamentalPhysics}.  Their value is largely attributable to their long trapping times which allow rapid accumulation of statistics, and since inhomogeneous broadening effects are very small, linewidths can be very narrow \cite{Ca1, Ca2, Ca3}.  Barium is of particular interest because of its potential for experimental observation of atomic parity nonconservation\cite{Fortson}.  To date, comprehensive studies of the lifetimes of metastable states \cite{lifetime}, hyperfine parameters of the odd isotopes $^{135}\mathrm{Ba}^+$ and $^{137}\mathrm{Ba}^+$ and isotope shifts \cite{Arroe} and branching ratios \cite{Kurz} have all been performed.  Also, barium provides a means by which to test models of nuclear structure, in particular a comparison between calculated and measured higher order nuclear moments\cite{octupole}.  Ab initio calculations are often not possible and require experimental input of atomic parameters.  We present the first precision measurement of the Land\'{e} g factor of the $5\mathrm{D}_{5/2}$ state of BaII, which we find to differ significantly from previously the accepted value of 1.12, derived from astronomical observation\cite{BMoore, Back, Curry, CMoore}.  Along with values for the g factors of the $6\mathrm{S}_{1/2}$,  $6\mathrm{P}_{1/2}$, $6\mathrm{P}_{3/2}$ and $5\mathrm{D}_{3/2}$, this completes the measurement of g factors for all low-lying (infrared and visible) atomic states of BaII\cite{Poulson, Knoll}.

For this experiment we perform high precision optical spectroscopy on single trapped $^{138}\mathrm{Ba}^+$ ions.  Neutral barium is photoionized by a two-step process; first, we excite an intercombination transition at 791 nm with an external cavity diode laser (ECDL) and then subsequently ionize these excited atoms with pulses from a nitrogen laser at 337 nm.  The first step allows for isotope selection\cite{Steele}.  The ions are confined in a radio frequency (RF) linear trap with approximately 1 W of RF  at a frequency $\Omega_{trap}\approx 2\pi\times12.38$ MHz and DC endcap bias of 670 V.  We measure trap secular frequencies in the radial and axial directions to be $\left(\omega_x, \omega_y, \omega_z\right)\approx 2\pi\times\left(2.5, 1.4, 0.6\right)$ MHz from the secular motion sidebands of measured spectra.  Current-carrying coils create a magnetic field of approximately 0.2-0.5 mT perpendicular to the k-vector of the cooling beam to prevent optical pumping into degenerate dark states and to separate Zeeman levels for the experiment.  Ba$^+$ has a lambda structure with a strong dipole $6\mathrm{S}_{1/2}\rightarrow 6\mathrm{P}_{1/2}$ transition at 493.4 nm used for Doppler cooling as shown in figure 1.  This has a branching ratio of approximately 25\% to the metastable $5\mathrm{D}_{3/2}$ state and hence requires a repump laser at 649.7 nm to depopulate that state.  A home-built ECDL at 986 nm is frequency doubled in a KNbO$_3$ bow-tie cavity to provide 493 nm light.  Approximately 10 $\mu$W is sufficient for cooling.  A second ECDL provides 649.7 nm light, of which about 50 $\mu$W is sent to the ion for the repump.  The two beams are sent to the ion via optical fiber to ensure perfect overlap, ease of alignment and ideal spatial mode with approximately 50 $\mu$m focus for low background scatter.  

\begin{figure}[H]
\centering
\includegraphics[scale=0.250]{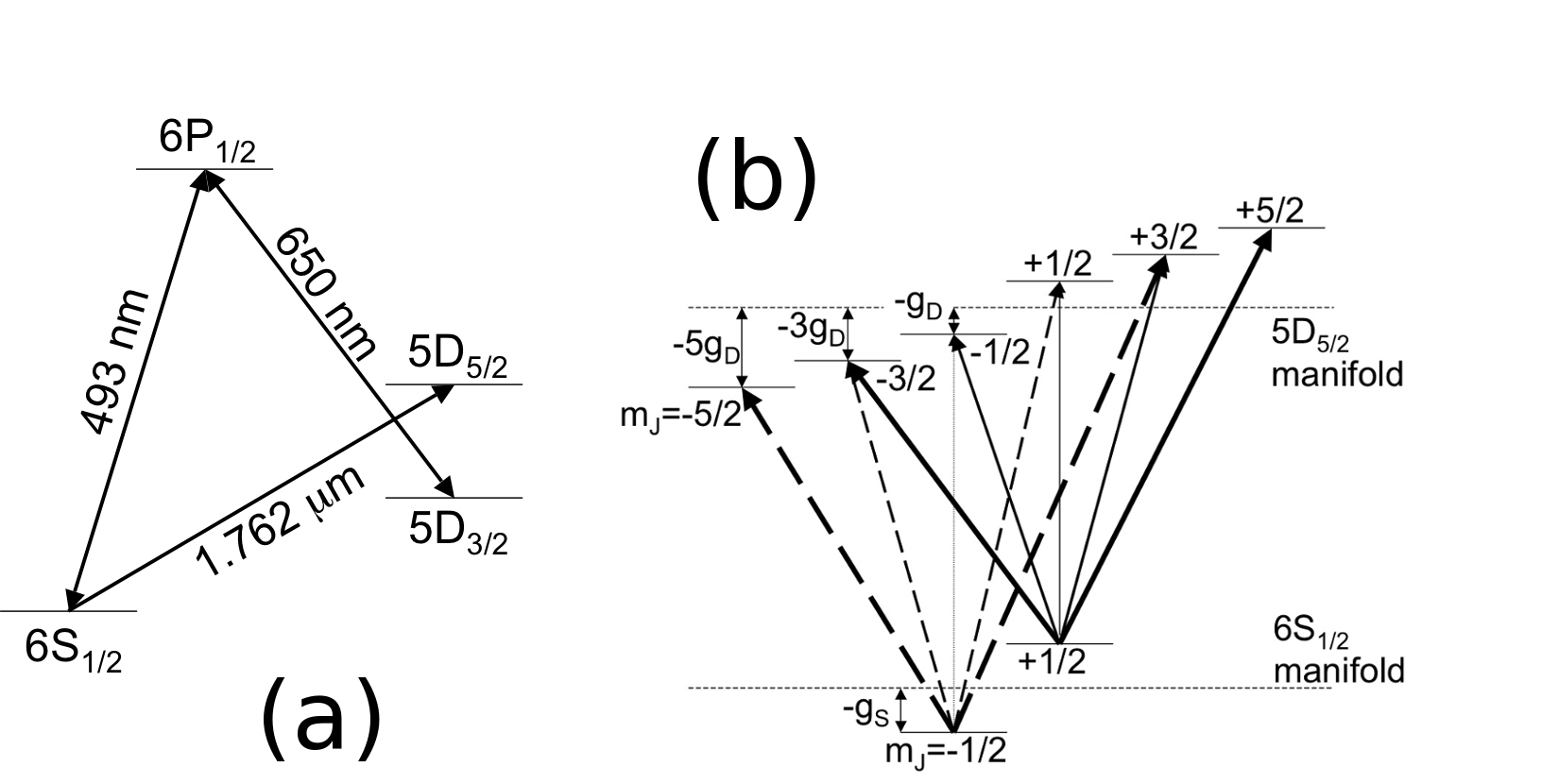}
\caption{(a) Fine structure of $^{138}\mathrm{Ba}^+$, showing the relevant states and transition wavelengths.  The 493 nm transition is used for cooling and fluorescence detection and is provided by a frequency-doubled ECDL.  A second ECDL at 650 nm pumps out of the $5\mathrm{D}_{3/2}$ state.  The 1.762 $\mu$m laser is a Tm-doped fiber laser.  Lifetimes in the D states are 30 s and 79 s for $5\mathrm{D}_{5/2}$ and $5\mathrm{D}_{3/2}$, respectively.  (b) Relative shifts of the observed transitions between ground and excited state Zeeman levels in $^{138}\mathrm{Ba}^+$, not to scale.  Allowed transitions from the $m_J=+1/2$ and $m_J=-1/2$ Zeeman levels are represented by solid and dashed arrows, respectively, with the relative strengths indicated by the intensity of the line.  $\Delta m_J=0$ transitions are suppressed by the relative orientation of the k-vector of the light to the quadrupole radiation pattern and were never observed.  Frequency shifts are indicated in units of $\frac{1}{2\hbar}\mu_B B$ for Zeeman levels with negative spin.  Those with corresponding positive spin shift upward by an equal amount.}
\end{figure}

We determine the relative frequencies of transitions between the Zeeman levels of the $6\mathrm{S}_{1/2}$ and $5\mathrm{D}_{5/2}$ states via excitation with a narrowband 1.762 $\mu$m thulium-doped fiber laser (Koheras Adjustik$^\mathrm{TM}$).  The laser is frequency stabilized via Pound-Drever-Hall (PDH) lock to a temperature-stabilized Zerodur$^\mathrm{TM}$ cavity with a finesse of approximately 1000 and free spectral range of 0.5 GHz placed in high vacuum\cite{600Hz}.  Coherence times of approximately 150 $\mu$s of Rabi oscillations between the ground and excited state indicate a laser linewidth of no greater than 10 kHz\cite{Dietrich}.  A double-pass acousto-optic modulator (AOM) shifts the frequency of a part of the laser power and introduces the frequency modulation necessary for a locking signal.  A second AOM in the beam path to the trap shifts the frequency of the remaining beam with respect to the cavity and can be scanned over a range of 80-120 MHz to observe transitions of the ion.  Exciting the ion to the $5\mathrm{D}_{5/2}$ state (lifetime $\tau\approx$ 30 s) takes the ion out of the cooling cycles and into a ``dark" state.  State detection is performed by monitoring the ion fluorescence at the 493 nm cooling laser transition, which is collected with an NA=0.28 microscope objective, spatially filtered and sent to a photomultiplier tube (PMT).  With photon counts of approximately 3000/s while fluorescing, we are able to distinguish bright and dark states with over 99\% efficiency after 10 ms observation time \cite{Dietrich}.

A small portion (approximately 1 $\mu$W) of the 493 nm light is split off from the cooling beam, polarized, sent through a quarter-wave plate and focussed to a spot size of approximately 70 $\mu$m onto the ion along the magnetic field direction.  This beam has purely circular polarization and optically pumps the ion into the $m_j=-1/2$ Zeeman sublevel of the $6\mathrm{S}_{1/2}$ state.  Fine-tuning of the relative direction of the field at the ion and the optical pumping beam is accomplished by adjusting the currents in two compensation coils orthogonal to each other and the main coil.  Optical pumping efficiency is better than 90\%.  Without the optical pumping, the ion population is in an approximately equal mixture of ground state Zeeman levels prior to excitation with 1.762 $\mu$m light.  Optical pumping will enhance or extinguish transitions in the ion's spectrum depending on the ground state level from which the transition begins.  By this method, we are able to identify specific transitions in the spectrum.

The sequence of the experiment is then as follows.  The ion is cooled with 493 and 650 nm light.  The 493 nm cooling light is switched off to prevent power broadening of the ground state while driving the $6\mathrm{S}_{1/2}\leftrightarrow 5\mathrm{D}_{5/2}$ transition.  Delaying turning off the 650 nm light pumps the ion to the ground state, during which time optical pumping light may or may not be applied to reach a definite Zeeman level of the ground state.  The 1.762 $\mu$m light is then switched on for 20 $\mu$s to drive the $6\mathrm{S}_{1/2}\rightarrow 5\mathrm{D}_{5/2}$ transition.  This pulse is line-triggered to reduce fluctuation due to magnetic perturbations.  The duration of the pulse is chosen to be approximately a $\pi$ Rabi rotation between the ground and excited state for carrier transitions and lower order micromotion sidebands, while suppressing the excitation of weaker sideband transitions.  After the pulse is applied, the cooling lasers are switched on and the ion's state is determined by observing its fluorescence on the PMT.  The sequence is repeated 40 to 100 times for each value of the shift AOM to accumulate statistics.  Shelving events are counted and the frequency of the laser shifted by the single-pass AOM to obtain the full spectrum.  We perform the cycle at six different values of the magnetic field and track the shift of features in the spectrum.  The entire dataset is shown in Fig 2(a), with a representative spectrum for the main coil current of 2.1 Amps in Fig 2(b).

\begin{figure}[H]
\centering
\includegraphics[scale=0.14]{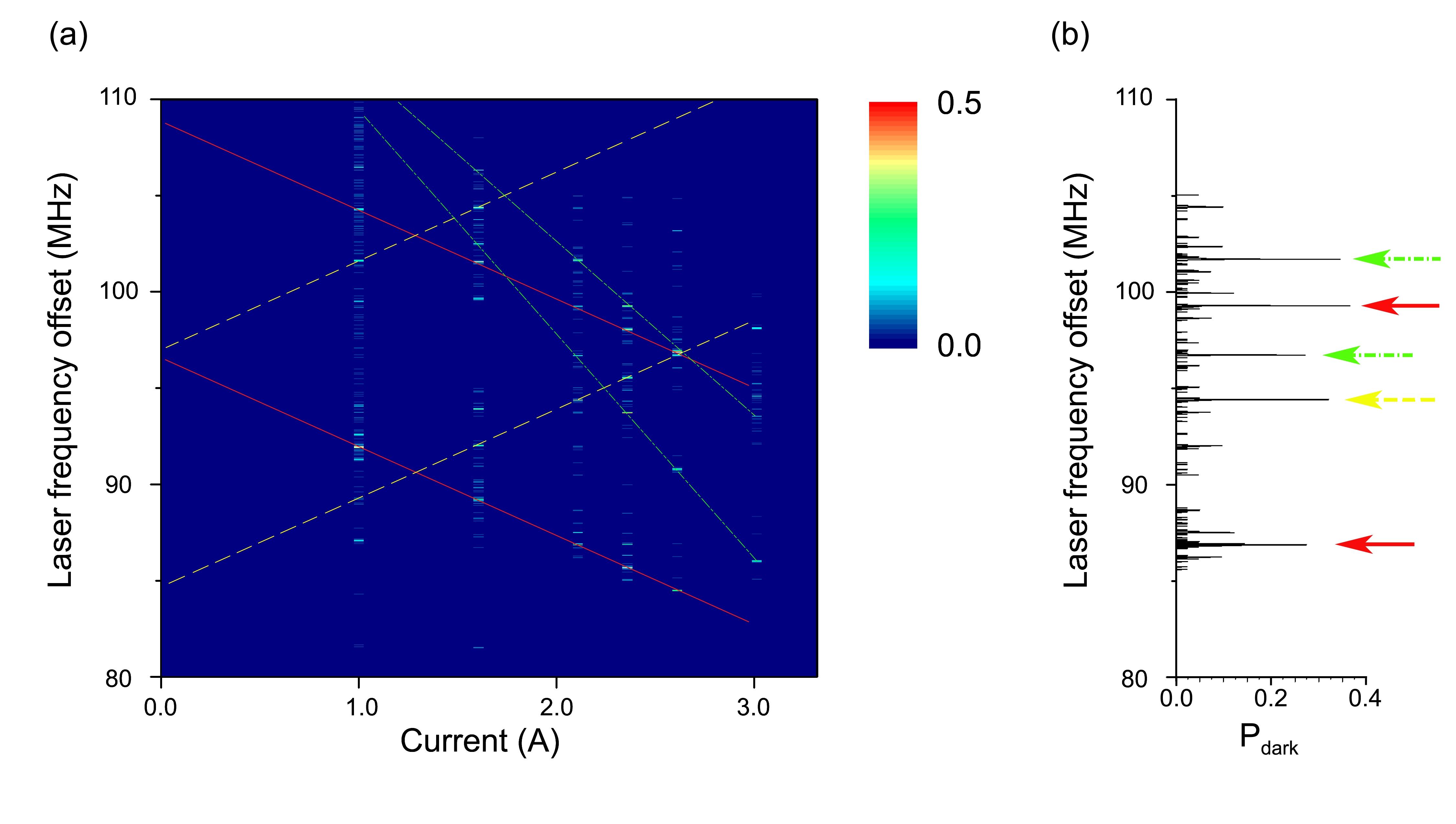}
\caption{(a) All spectra measured without optical pumping showing $6\mathrm{S}_{1/2}\leftrightarrow 5\mathrm{D}_{5/2}$ transitions at six different values for the main coil current.  The lines on the figure are inserted by hand to guide the eye.  Resonances not crossed by the guiding lines were identified as secular sidebands by their characteristic frequency separation from main transitions.  The dashed (yellow) lines are the $\Delta m=+1$ from $m_J=+1/2$ and its 1$^\mathrm{st}$ order micromotion sideband, the solid (red) lines are the $\Delta m=-1$ from $m_J=-1/2$ and its 1$^\mathrm{st}$ order micromotion sideband and dash-dot (green) the 2$^\mathrm{nd}$ order sidebands of $\Delta m=-2$ from $m_J=-1/2$ and $\Delta m=-1$ from $m_J=+1/2$.  Secular motion sidebands at 2.5, 1.4 and 0.6 MHz are visible as well but deliberately not emphasized by added lines (b) A representative spectrum taken without optical pumping at 2.1 Amps.  Arrow line styles (colors) correspond to the same scheme used to label transitions in (a).  \it color online\rm}
\end{figure}

We obtain a spectrum of peaks with typical widths of approximately 50 kHz that shift monotonically with changing field coil current.  Peak heights near the central frequency of the single-pass AOM are 30-50\% dark state probability at each frequency step without optical pumping as is to be expected because the two ground state Zeeman levels are equally occupied.  Optical pumping for 200 $\mu$s prior to sending the 1.762 $\mu$m laser pulse extinguishes transitions that originate from the $m_J=+1/2$ Zeeman sublevel, while increasing the peak heights to over 90\% dark state probability for those which originate from the $m_J=-1/2$ level.  We differentiate micromotion sideband transitions from carrier transitions by their relative strengths; carrier Rabi oscillations between the ground state and excited state have typical $\pi$ pulse times of 10 $\mu$s while those on sidebands are two to three times as long.  


We clearly observe a carrier transition and its first micromotion sideband, which consistently shift parallel to one another with changing current as indicated in figure 2(a).  Using these lines we can precisely determine the trap RF frequency $\Omega_{trap}$ to be $12.3759\pm 0.0025$ MHz, in excellent agreement with direct measurement of this frequency.  A second pair of lines which exhibits the same characteristics but whose transition frequency increases with increasing field was also observed.  Optical pumping enhances the first pair but extinguishes the second.  We thus conclude that these lines originate from ground state Zeeman levels of opposite spin.  Two weaker lines which converge to a frequency approximately 24.6 MHz higher at zero coil current than the other two measured carriers and display opposite behavior when optically pumped were observed as shown in figure 2(a).  These correspond to second order micromotion sidebands, since their zero current frequency is higher than the zero current frequency of observed carrier transitions by an amount equal to exactly twice the trap frequency.

The error in bright/dark counts is taken to be binomial and the spectral features are fit with a Lorentzian to obtain the precise central frequency of transitions.  The statistical uncertainty in peak positions is typically one to several kHz.  Because the $6\mathrm{S}_{1/2}\rightarrow 5\mathrm{D}_{5/2}$ transition is an electric quadrupole transition, $\Delta$m=0, 1 or 2 transitions are all in principle allowed by selection rules.  However, because the k-vector of the 1.762 $\mu$m laser is perpendicular to the quantization axis set by the B-field, the radiation pattern for $\Delta m=0$ transitions suppresses those features and they were never observed in the measured spectra.  The polarization of the 1.762 $\mu$m laser and radiation patterns allow both $\Delta m=1$ and 2 transitions, with the latter being slightly stronger \cite{Roos}.  Expected carrier transitions and relative shifts are shown in figure 1(b).


By subtracting the measured trap frequency from first order micromotion sidebands and twice that frequency from second order sideband positions, we calculate the position of the carrier frequency for four Zeeman transitions: $\Delta m=+1$ from $\left|6\mathrm{S}_{1/2}\right.$; $\left.m_J=+1/2\right>$, $\Delta m=-1$ from $\left|6\mathrm{S}_{1/2}\right.$; $\left.m_J=-1/2\right>$, $\Delta m=-1$ from $\left|6\mathrm{S}_{1/2}\right.$; $\left.m_J=+1/2\right>$ and $\Delta m=-2$ from $\left|6\mathrm{S}_{1/2}\right.$; $\left.m_J=-1/2\right>$ (see Fig. 3).  The other Zeeman transitions fall well outside of the available bandwidth of the AOM.  The Zeeman levels of the D state shift by $g_Dm_J\mu_B B$ and those of the ground state by $g_Sm_J\mu_B B$, in an applied magnetic field $B$, with $\mu_B=1.40$ MHz$\cdot$G$^{-1}$ the Bohr magneton.  The predicted shift directions and relative magnitudes are summarized in Fig. 1(b).  Taking the magnetic field and zero field frequency to be unknown parameters and eliminating their dependence by strictly using frequency differences among measured lines at individual values for the magnetic field, we algebraically solve for the ratio of the g factors of the two states.  The g factor ratio as calculated from the four different measured lines agree to within the statistical error, as do the value for the ratio at different coil current.  We are further able to solve for the applied field and find that we apply 0.067 mT of magnetic field for each amp of coil current, to a maximum of 0.2 mT at the 3.0 A value used here, well within the linear Zeeman regime.  Inputting the accepted value for $g_{6\mathrm{S}_{1/2}}=2.0024906(11)$\cite{Knoll}, we are able to solve for $g_{5\mathrm{D}_{5/2}}=1.2020\pm0.0005_{stat}$.  Strict LS coupling predicts a g factor of 6/5=1.2 for the $5\mathrm{D}_{5/2}$ state.  Leading order corrections come from relativistic effects, interactions between the valence electron and the core shell and QED shifts from electron self-energy and vacuum polarization.  The largest of these occurs at the $\alpha(\alpha Z)^2$ level, approximately the level at which we observe a deviation from 6/5\cite{Khetselius, QED}.

\begin{figure}[H]
\centering
\includegraphics[scale=0.65]{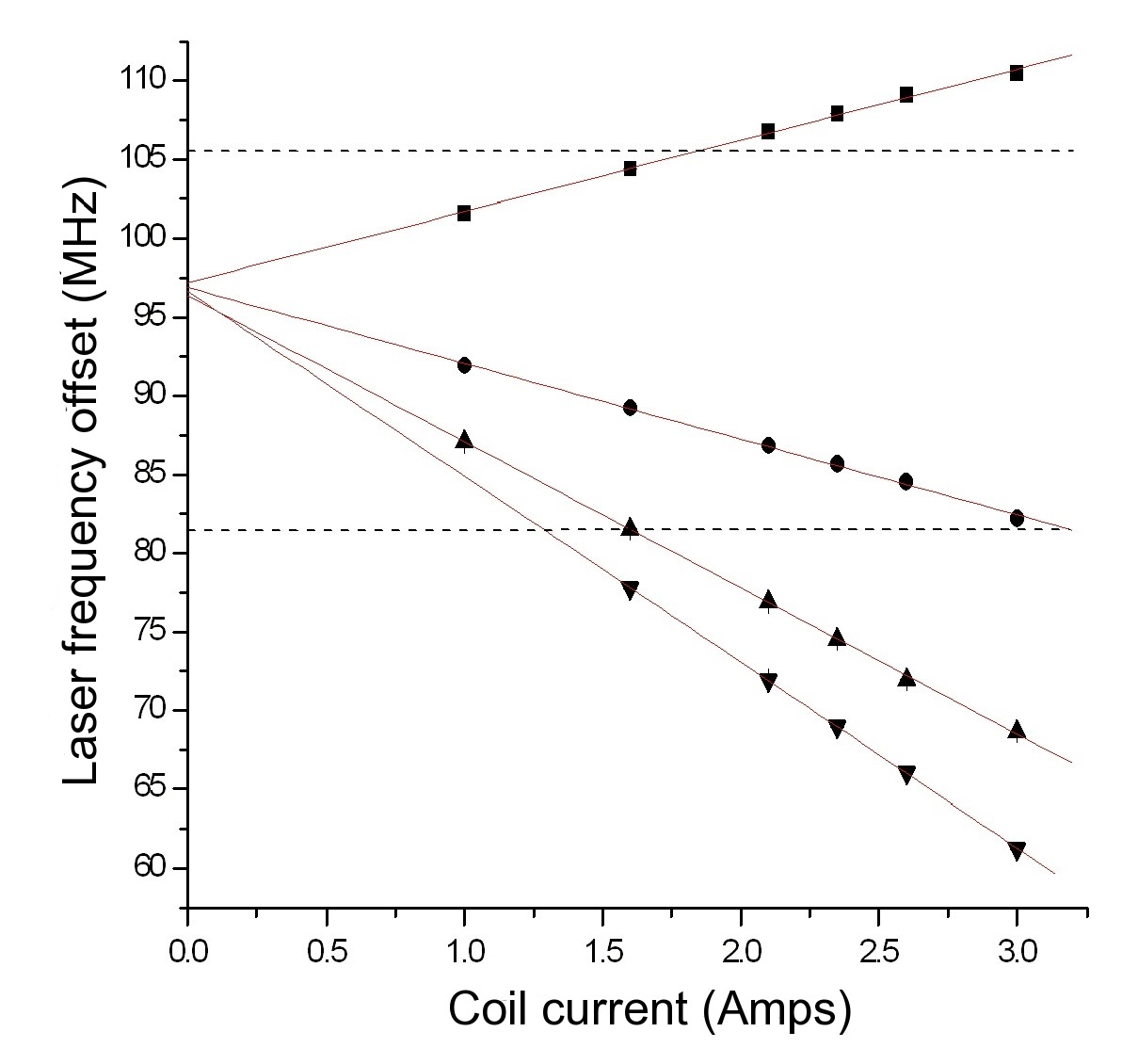}
\caption{Measured and extrapolated positions of carrier transitions.  Squares denote $\Delta m=+1$ from $m=+1/2$, circles $\Delta m=-1$ from $m=-1/2$, triangles $\Delta m=-1$ from $m=+1/2$ and upside-down triangles $\Delta m=-2$ from $m=-1/2$.  Below 82 MHz and above 105 MHz (horizontal dashed lines in the figure), positions of carrier transitions were extrapolated from the position of micromotion sidebands due to insufficient bandwidth of the single-pass AOM to reach frequencies out of that range.  The micromotion frequency itself was determined by the difference between measured carrier and first order sideband positions of the $\Delta m=-1$ from the $m_J=+1/2$ state.  Fits are to the function $f(I)=\alpha\sqrt{I^2+B_o^2}+f_o$, where $I$ is the main coil current and $B_o$ is a small offset magnetic field at zero coil current found to be less than 0.1\% of the magnetic field from the main current coils and $f_o$ is the transition frequency between $6\mathrm{S}_{1/2}$ and $5\mathrm{D}_{5/2}$ at zero field.  \it color online\rm}
\end{figure}

We suspect several sources of systematic error.  Because we are not concerned with linewidths or transition strengths, we consider only effects that influence the relative frequencies of features in an individual run.  Two sources of error related to trapping parameters are the variation of the ion's position with respect to the magnetic field and AC Zeeman shifts caused by currents induced in the trap by the high RF voltages.  We estimate the former to be of the level of $10^{-4}$ as the ion's position varies by much less than a few microns over the course of the experiment, as observed on a electron-multiplied CCD camera, and magnetic gradients are of the order 40 nT/$\mu$m.  AC Zeeman shifts have been seen to affect g factor measurements, albeit at the $10^{-4}$ level in slightly different trap geometries\cite{Sherman}, but these are highly suppressed in our trap design because of the large distance between the ion and the nearest grounded electrode through which current is able to pass to ground.  Several potential sources of frequency error are related to the shelving laser itself.  Drifts can occur both on the locking cavity due to pressure or temperature variations and in the phase of the 60 Hz AC voltage to which the beginning of the shelving pulse is triggered.  This is only a problem during the course of a single run at each individual magnetic field, since the extraction of the g factor comes from each field value rather than a fit as a function of field.  We can use the observed change in the position of the $6\mathrm{S}_{1/2}\rightarrow 5\mathrm{D}_{5/2}$ transition as a combined measure of these effects.  We extract the central (non Zeeman-shifted) value for the $6\mathrm{S}_{1/2}\rightarrow 5\mathrm{D}_{5/2}$ transition, which lies midway between the $\Delta m=+1$ from $\left|6\mathrm{S}_{1/2}\right.$; $\left.m_J=-1/2\right>$ and $\Delta m=-1$ from $\left|6\mathrm{S}_{1/2}\right.$; $\left.m_J=+1/2\right>$ transitions.  This value agrees to within 4 kHz for the different values of coil current.  From this uncertainty we estimate that the shift over the course of an individual run, which is at most one hour in duration, is at worst a few kHz.  This represents the dominant contribution to the systematic uncertainty and we can safely assume the overall systematic error in the g factor measurement to be $\pm 0.0001$.

In conclusion, we have measured the Land\'{e} g factor of the $5\mathrm{D}_{5/2}$ state of the Ba$^+$ ion to be $1.2020\pm0.0005_{stat}\pm0.0001_{sys}$, a significant improvement over the uncertainty of previously published values.  We find that it differs significantly from previous measurements.  With its importance for experiments testing nuclear structure, parity nonconservation and for applications as a frequency standard, barium's atomic structure is of continued interest, both theoretically and for future experimental study.

The authors would like to acknowledge the contributions of Jennifer Porter, Joanna Salacka, Aaron Avril and Tom Chartrand to the work.  Research support was provided by National Science Foundation Grants Nos. 0758025 and 0904004.


\begin{thebibliography}{99}
\bibitem{Prestage}J. D. Prestage, R. L. Tjoelker \& L. Maleki, Phys. Rev. Lett. \bf 74\rm, 3511-3514 (1995).
\bibitem{Fortier}T. M. Fortier, et. al., Phys. Rev. Lett. \bf 98\rm, 070801 (2007).
\bibitem{Dzuba}V. A. Dzuba \& V. V. Flambaum, Phys. Rev. A \bf 61\rm, 034502 (2000).
\bibitem{standards}A. A. Madej \& J. E. Bernard, ``Single-Ion Optical Frequency Standards and Measurement of their Absolute Optical Frequency" in \it Frequency Measurement and Control\rm, (Springer Berlin, Heidelberg 2001).
\bibitem{clocks}J. Sherman, W. Trimble, S. Metz., W. Nagourney \& E. N. Fortson, \it Progress on Indium and Barium single-ion optical frequency standards\rm, 2005 Digest of the LEOS Summer Topical Meetings, IEEE Cat. No. 05TH8797 (IEEE, New York, 2005).
\bibitem{FundamentalPhysics}V. A. Dzuba \& J. S. M. Ginges, Phys. Rev. A \bf 73\rm, 032503 (2006).
\bibitem{Ca1}A. Kreuter, et. al., Phys. Rev A \bf 71\rm, 032504 (2005).
\bibitem{Ca2}J. Benhelm, G. Kirchmair, U. Rapol, T. K\"{o}rber, C. F. Roos \& R. Blatt, Phys. Rev. A \bf 75\rm, 032506 (2007).
\bibitem{Ca3}P. A. Barton, C. J. S. Donald, D. M. Lucas, D. A. Stevens, A. M. Steane \& D. N. Stacey Phys. Rev. A \bf 62\rm, 032503 (2000).
\bibitem{Fortson}E. N. Fortson, Phys. Rev. Lett. \bf 70\rm, 2383-2386 (1993).
\bibitem{lifetime} A. A. Madej \& J. D. Sankey, Phys. Rev. A \bf 4\rm, 2621 (1990). 
\bibitem{Arroe} O. H. Arroe, Phys. Rev. \bf 79\rm, 836Ð838 (1950).
\bibitem{Kurz}N. Kurz, M. R. Dietrich, Gang Shu, R. Bowler, J. Salacka, V. Mirgon \& B. B. Blinov, Phys. Rev. A \bf 77\rm, 060501(R) (2008).
\bibitem{octupole}K. Beloy, A. Derevianko, V. A. Dzuba, G. T. Howell, B. B. Blinov \& E. N. Fortson, Phys. Rev. A \bf 77\rm, 052503 (2008).
\bibitem{BMoore}B. E. Moore, Ann. der Phys. [4] \bf 25\rm, 314 (1908).
\bibitem{Back}E. Back, Ann. der Phys. [4] \bf 70\rm, 360-3 (1923).
\bibitem{Curry}J. J. Curry, Journal Phys. Chem. Ref. Data \bf 33\rm, 725 (2004).
\bibitem{CMoore}C. E. Moore, \it Atomic Energy Levels v. 3\rm, Nat. Bur. Std. Circ. No. 467 (US Gov't. Printing Office, Washington, D.C. 1958).
\bibitem{Poulson} O. Poulson \& P. S. Ramanujan, Phys. Rev. A \bf 14\rm(4) 1463-7 (1976).
\bibitem{Knoll} K. H. Kn\"{o}ll, G. Marx, K. H\"{u}bner, F. Schweikert, S. Stahl, Ch. Weber \& G. Werth, Phys. Rev. A \bf 54\rm(2) 1199-1205 (1996).
\bibitem{Steele}A. V. Steele, L. R. Churchill, P. F. Griffin \& M. S. Chapman, Phys. Rev. A \bf 75\rm, 053404 (2007).
\bibitem{600Hz}N. Yu, X. Zhao, H. G. Dehmelt \& W. Nagourney, Phys. Rev A \bf 50\rm, 2738-2741 (1994).
\bibitem{Dietrich}M. R. Dietrich, N. Kurz, T. Noel, G. Shu \& B. B. Blinov, Phys. Rev. A \bf 81\rm, 052328 (2010).
\bibitem{Roos} C. Roos, Ph. D. thesis, University of Innsbruck, 2000, Downloadable at http://heart-c704.uibk.ac.at/publications/dissertation/roos\textunderscore diss.pdf
\bibitem{Khetselius} O. Yu Khetselius, Phys. Scr. T\bf 135\rm, 014023-6 (2009).
\bibitem{QED}D. A. Glazov, et. al., Phys. Rev. A \bf 70\rm, 062104 (2004).
\bibitem{Sherman}J. A. Sherman, A. Andalkar, W. Nagourney \& E. N. Fortson, Phys. Rev. A. \bf 78\rm, 052514 (2008).

\end{thebibliography}
\end{document}